\documentstyle[aps,multicol]{revtex}
\begin{document}
\title{Rotating Superconductors and\\ the Frame-independent London Equation}
\author{Mario Liu~\cite{email}}
\address{Institut f\"ur Theoretische Physik, Universit\"at Hannover,
              30167 Hannover, Germany, EC}
\maketitle
\begin{abstract}
A frame-independent, thermodynamically exact London equation is presented, 
which is especially valid for rotating superconductors. A direct result is 
the unexpectedly high accuracy ($\sim10^{-10}$) for the usual expression of 
the London moment. \end{abstract}
\draft
\begin{multicols}{2}

The defining property of superconductors to expel any magnetic field from 
its bulk reverses when it is rotated, and a spontaneous field (or London 
moment) 

\begin{equation}\label{0}
B=-2\Omega/\gamma, \quad\gamma\equiv e/mc, 
\end{equation} 
appears, where $\Omega$ is the rotational velocity, while 
$m,\, e(<0)$ are respectively the bare mass and charge of the 
electron~\cite{London,Cabr}. (In MKSA units, $\gamma=e/m$.) This is 
intriguing because some exact cancellations must have taken place before 
microscopic parameters such as $e$ or $m$ will appear on a macroscopic 
level, in a complicated many body system that any superconductor is. So in 
this sense, the London moment may perhaps be compared to the quantum Hall 
effect~\cite{laf}. The bare charge $e$ holds its place because the 
superfluid velocity, 

\begin{equation}\label{1}
\mbox{\boldmath$v$}_s=(\hbar/2m)\nabla\varphi -\gamma{\bf A}, 
\end{equation}
must be gauge-invariant. (${\bf A}$ is the vector potential, and $\varphi$ 
the macroscopic phase variable characterizing the spontaneously broken 
symmetry.) Although this type of arguments does not apply to $m$, a common 
prefactor in Eq(\ref{1}), many authors have convincingly argued why 
nevertheless, and to what extent, $m$ remains unrenormalized in the London 
moment~\cite{Cabr,PWA,Bra,baym}. 

In this paper, a symmetry argument as simple as the gauge invariance is 
invoked to show why it is the bare mass that enters Eqs(\ref{0},\ref{1}). 
The basic point is, $v_s$ is only as defined, with the bare mass $m$, a 
velocity under Galilean transformations.

There is a preferred coordinate system in superconductors, that of the 
lattice, and most useful properties are best understood in this system. 
Confining one's interest to these, the transformation property of $v_s$ is 
irrelevant, and de Gennes' famously pointed statement,   that the mass $m$ 
is completely arbitrary, and might well be taken as that of the 
sun~\cite{dG}, valid.  

Sometimes, however, other coordinate systems are more advantageous, 
especially if the lattice itself (when rotating or subject to a sound field) 
is not an inertial system. A theory that consistently relates the results 
from different frames is then desirable. And how $v_s$ transforms is an 
important input for such a frame-independent theory.  As will be shown 
below, the London moment may be understood as a direct consequence of $v_s$ 
being a velocity. 

The superfluid velocity $v_s$ is a concept perhaps more of superfluid 
helium, and less universally employed in the context of superconductors. The 
reason frequently given is again the presence of the lattice, which breaks 
the translational invariance for the electrons. (Even without the lattice, a 
uniform background charge breaks the Galilean symmetry.) This is not 
convincing, as the transformations may well be taken to include the ions. 
Then the usual invariance under translations, rotations and Galilean boosts 
is restored, and we retain local conservation of (total) momentum and 
angular momentum. So, as in superfluids, $v_s$ is a perfectly legitimate 
thermodynamic variable in superconductors. It characterizes the 
spontaneously broken phase symmetry, and (as we shall see) gives rise to 
equilibrium mass and charge current simultaneously.  

In contrast to the exact gauge symmetry, the Galilean transformation 
property of $v_s$, and hence Eq(\ref{0}), are approximative and subject to 
relativistic corrections. This has been carefully considered by various 
authors, obtaining values~\cite{Cabr} that lie between $10^{-4}$ and 
$10^{-5}$. The relevant quantity here is the work function $W$ of the 
electron, or the electrochemical potential, over $mc^2$. As pointed out by 
these authors, this value is subject to further scrutiny, as possibly 
important effects such as the centrifugal force of the rotating 
frame~\cite{Bra}, or the presence of the symmetry breaking 
lattice~\cite{baym}, have not been included. 

Eliminating these sources of uncertainties, especially by including the 
ions explicitly, the accuracy is shown in this paper to be given by 
$\mu_{\rm chem}/c^2$, where $\mu_{\rm chem}$ is the chemical potential of 
the superconductor, or the energy per unit mass of adding a (neutral) atom 
to the crystal. Due to the much greater mass of the atoms, $\mu_{\rm 
chem}/c^2$ is tiny, of order $10^{-10}$. This makes the correction 
negligible for all practical purposes, and renders Eq(\ref{0}) about three 
orders of magnitude more precise than any present experimental technique to 
determine the electron mass directly~\cite{d}. 

To understand the connection between Eq(\ref{0}) and the Galilean 
transformation property of $v_s$, consider the London equation, which for 
more than half a century has been very useful to account for static 
properties of superconductors, especially the Mei\ss ner effect and the 
associated magnetic healing phenomena. It is obtained by taking ${\bf j}_s= 
\varrho_s \mbox{\boldmath$v$}_s$ in the static Maxwell equation 
$\nabla\times{\bf H}= \gamma{\bf j}_s$, where $\varrho_s$ is the superfluid 
density, a stiffness constant. (The superconducting number density 
$n_s\equiv\varrho_s/m$ is probably more familiar.) Applying a further curl 
on both sides, we arrive at (what we for simplicity shall here call) the 
London equation, $\nabla\times\nabla\times {\bf  H} = -\varrho_s\gamma^2{\bf 
B}$, (${\bf B}\equiv\nabla\times{\bf A}$,) which yields the magnetic 
penetration depth $\lambda=\pm\gamma\sqrt{\varrho_s}$ (for $H=B$)~\cite{HL}. 
Unfortunately, the same equation also insists on vanishing $B$-fields in the 
bulk, irrespective of a possible rotation $\Omega\not=0$. 

Because of its curious ineptness to describe rotating superconductors, the 
London equation is not usually consulted in this context, (a notable 
exception is~\cite{Cabr}). Instead, Eq(\ref{1}) is considered alone, with a 
large dose of healthy intuition to compensate for its incompleteness: {\em 
If we rotate a bulk cylinder of superconductor, it is essential that over 
the sample as a whole $v_s=v_{\rm lattice}$, because otherwise very large 
currents would flow\dots}~\cite{PWA}. This appears convincing, as taking the 
curl of Eq(\ref{1}) with $v_s=v_{\rm lattice}$ yields Eq(\ref{0}) with 
$\Omega=\frac{1}{2}\nabla\times v_{\rm lattice}$, but it is not really 
consistent: Inserting $v_s=v_{\rm lattice}$ also in ${\bf j}_s= \varrho_s 
\mbox{\boldmath$v$}_s$ leads exactly to the large electric current that was 
to be avoided in the first place. And setting $v_s=v_{\rm lattice}$ while 
considering linear motions of the superconductor plays further havoc with 
physics. 

This paper reports the thermodynamic derivation of a generalized London 
equation that is (i) valid in an arbitrary inertial frame, and (ii) given in 
terms of the macroscopic, measured fields. It is 

\begin{equation}\label{2}
\nabla\times\nabla\times{\bf H}_0= -\varrho_s\gamma(\gamma{\bf B}/\eta+ 
2\mbox{\boldmath$\Omega$}), \end{equation}
where ${\bf H}_0$ denotes the field in the local rest frame of the lattice, 
and 

\begin{equation}
\eta\equiv1+ \mu_{\rm chem}/c^2
\end{equation}
is the relativistic correction. Clearly, this equation accounts for the 
bulk value of the field,  $B=-2\eta\Omega/\gamma$, and the attendant 
magnetic healing, both with and without rotations. 

Eq(\ref{2}) is obtained by combining 

\begin{eqnarray}\label{3}
\nabla\times{\bf H}_0=\gamma{\bf j}_s,\\
{\bf j}_s=\varrho_s(\mbox{\boldmath$v$}_s/\eta - \mbox{\boldmath$v$}_n). 
\label{4}\end{eqnarray}
Both will be derived carefully below, but let us first establish a 
qualitative understanding: Eq(\ref{3}) is an Euler equation, expressing the 
fact that the free energy is minimal with respect to variations in the 
vector potential $\delta A$; Eq(\ref{4}) is a constitutive relation, 
expressing the thermodynamic conjugate variable ${\bf 
j}_s\equiv\partial\varepsilon/\partial \mbox{\boldmath$v$}_s$ as a function 
of the two velocities, $v_s$ and $v_n$, where $v_n$ is the velocity of the 
lattice points. As the left side of Eq(\ref{3}) is invariant under a 
Galilean transformation, so must the right side $\sim j_s$ be, which may 
therefore only depend on the velocity difference. Eqs(\ref{4}) states this 
fact to linear order, including the relativistic correction. (This is where 
the information that $v_s$ is a velocity enters the equations.)

To understand the transformation behavior of $v_s$, we start from the rest 
frame form of the Josephson equation, 

\begin{equation}\label{j}
(\hbar/2)\dot\varphi+ \mu_-=0. 
\end{equation}
Consider first the electrochemical potential $\mu_-$: Take the 
thermodynamic energy density $\varepsilon$ either as a function of  the mass 
and charge density, 

\begin{equation}\label{8}
{\rm d}\varepsilon=\mu\,{\rm d}\varrho+ \Phi\,{\rm d}\rho_e,
\end{equation}
 or as a function of the numbers of ions and electrons, 

\begin{equation}
{\rm d}\varepsilon=\mu_+\,{\rm d}n_+ + \mu_-\,{\rm d}n_-. 
\end{equation}
These two pairs of independent variables are related by 

\begin{equation}
\varrho=Mn_+ +mn_-, \qquad \rho_e=|e|(n_+ -n_-),
\end{equation}
with $M, m$ as their mass and $|e|, e$ as their charge, respectively.  As a 
result, the conjugate variables are related as 

\begin{equation}\label{10}
\mu_+=M\mu-e\Phi, \qquad\mu_-=m\mu+e\Phi.
\end{equation}
Note two points: (i) This is the only place where $m$ enters our 
considerations independently from Eq(\ref{1}), yet since $\varrho, \rho_e, 
n_+, n_-$ are all strictly conserved, $M, m, e$ are the bare parameters. 
(ii) Following the standard notation in  relativistic physics, we take the 
energy density  $\varepsilon$ to include the rest energy, $\varrho c^2$. 
Then all three chemical potentials contain a large constant term, especially 
 
\begin{equation}
\mu=\mu_{\rm chem}+c^2. 
\end{equation}
With the Josephson equation now given as 

\begin{equation}\label{6}
(\hbar/2m)\dot\varphi + \mu=-e\Phi/m,\quad 
\dot{\mbox{\boldmath$v$}}_s+\nabla\mu=e{\bf E}/m, 
\end{equation}
we may define a 4-vector 

\begin{equation}\label{4vec}
u_\alpha\equiv(\hbar/2m)\partial_\alpha\varphi -\gamma A_\alpha=(u_0, 
\mbox{\boldmath$v$}_s),
\end{equation}
such that a transformed $u' _\alpha$, in the system boosted by ${\bf v}$,  
is given as $(u'_0, \mbox{\boldmath$v$}'_s)=(u_0+{\bf 
v}\cdot\mbox{\boldmath$v$}_s/c, \mbox{\boldmath$v$}_s+u_0{\bf v}/c)$, or 
with Eq(\ref{6},\ref{4vec}), 

\begin{equation}
\mbox{\boldmath$v$}'_s=\mbox{\boldmath$v$}_s-\mu{\bf v}/c^2.
\end{equation} 
Taking $\mu/c^2=1+\mu_{\rm chem}/c^2\equiv\eta$, we see that 
$\mbox{\boldmath$v$}_s/\eta$ transforms as a velocity, rendering the 
combination in Eq(\ref{4}) invariant.

As is clear from Eq(\ref{8}), $\mu_{\rm chem}$ is the energy of adding an 
atom to the crystal at constant charge, divided by the mass of the atom. 
Estimating this energy as $10^4$K (boiling/melting temperature), or 
$10^{-19}$J, and the atom as having 50 proton mass $\approx10^5$ electron 
mass $\approx10^{-25}$kg,  we have $|\mu_{\rm chem}|\approx10^6$(m/s)$^2$, 
or $\eta\approx1-10^{-10}$.  

The main correction being so small, any inaccuracy in Eq(\ref{j}) will be 
even less important. For instance, the velocity dependent terms in the 
Josephson equation (not considered above as the rest frame form was chosen) 
are smaller if the macroscopic velocity stays below $\sqrt{\mu_{\rm 
chem}}\approx 10^3$m/s. The same is true for the centrifugal force, which 
slightly compresses the outer rim, and decompresses the center, of the 
superconductor. As a result, the density $\varrho$, and hence $\mu_{\rm 
chem}(\varrho)$, are  inhomogeneous. However, since $\mu_{\rm chem}-({\bf 
\Omega}\times\mbox{\boldmath$r$})^2/2$ remains constant~\cite{3cd}, this 
correction is again of order $(v/c)^2$. 

The reasons previous authors arrived at much larger relativistic 
corrections vary: Some considered the 4-vector 
$w_\alpha\equiv(\hbar/2m)\partial_\alpha\varphi$ that transforms as 
$\mbox{\boldmath$w$}'_s=\mbox{\boldmath$w$}_s-{\bf v}(\mu_-/mc^2)$. (See for 
instance Eq(10) of \cite{Bra}.) Mainly because of the smallness of the 
electron mass $m$, the associated relativistic correction $\mu_-/mc^2$ 
deviates more strongly from 1. Others did consider the 4-vector of 
Eq(\ref{4vec}), especially Cabrera~\cite{Cabr}, who in fact obtained the 
combination, $\mu_--e\Phi$ as the relevant factor in the boost 
transformation. The absence of ions in his considerations, however, 
prevented an interpretation of this combination as the chemical potential of 
the solid. In addition, the term $e\Phi$ was taken to be given only in the 
rotating, and not in the laboratory frame, and was therefore neglected.

We now proceed to prove Eqs(\ref{3}) by thermodynamic considerations, in a 
way that is similar to the minimization procedure of the Ginsburg-Landau 
functional~\cite{LL9}, though the procedure is executed in a general 
inertial frame here~\cite{nurmax}, not necessarily that of the lattice. The 
basic Gibbs relation is  

\begin{eqnarray}
{\rm d}\varepsilon&=& T\,{\rm d}s+ \mu\,{\rm d}\varrho+ 
\mbox{\boldmath$v$}_n\cdot {\rm d}\mbox{\boldmath$g$}^{\rm tot} 
+\sigma_{ij}{\rm d}\nabla_ju_i \nonumber\\ &&+{\bf E}_0\cdot{\rm d}{\bf D} 
+{\bf H}_0\cdot{\rm d}{\bf B} + {\bf j}_s\cdot{\rm d}\mbox{\boldmath$v$}_s. 
\label{5}\end{eqnarray}
The first four terms are that of a normal solid~\cite{mpp}, with 
$\varepsilon, s, \varrho, g^{\rm tot}$ denoting the densities of energy, 
entropy, mass and momentum, respectively, while $u_i$ is the displacement 
vector. The conjugate variables are defined by the respective derivatives, 
eg $T\equiv\partial\varepsilon/\partial s$, or $v_n\equiv\partial 
\varepsilon/\partial g^{\rm tot}$. The next two terms account for the 
presence of fields, where the subscript $_0$ denotes the respective field in 
the rest frame $\mbox{\boldmath$v$}_n =0$. The last term appears in 
superconducting or superfluid  phases, with the variable given by 
Eq(\ref{1}). Accepting that the energy is given as in Eq(\ref{5}), 
minimizing the free energy  while keeping the conserved quantities constant, 
stationarity with respect to variations in the vector potential,  
$$\delta\int (\varepsilon-Ts) =\int[{\bf H}_0\delta(\nabla\times{\bf 
A})-{\bf j}_s \delta(\gamma{\bf A})]+\dots$$ quickly leads to Eq(\ref{3}). 
The other equilibrium conditions are~\cite{3cd}: $\partial_i v^n_j+ 
\partial_j v^n_i=0$, $\partial_t\mbox{\boldmath$v$}^n+ 
\mbox{\boldmath$\nabla$} \mu=0$,
$\nabla_j\sigma_{ij}=0$, ${\bf E}_0=0$. 

Although the basic structure of Eq(\ref{5}) simply states the fact that the 
thermodynamic equilibrium of a superconducting crystal depends on the 
specified variables, there are perhaps three points that may seem puzzling 
at first: (i) Why is the term $\Phi\,{\rm d}\rho_e$ of Eq(\ref{8}) 
substituted by its partially integrated form, ${\bf E}_0\cdot{\rm d}{\bf D}$ 
of Eq(\ref{5})? (ii) Why do the two fields, ${\bf E}$ and ${\bf H}$, assume 
their local rest frame values? And (iii), to what extend is 
$\mbox{\boldmath$v$}_n$ the velocity of the lattice? The answer to the first 
question is connected to locality: ${\bf E}_0$ is a local function of ${\bf 
D}$, while $\rho_e$ from everywhere is needed to calculate $\Phi$. Local 
thermodynamics and hydrodynamics can only deal with quantities that preserve 
locality. The second question is discussed in details in~\cite{kh}, and is a 
result of the fact that the conserved, total momentum density $g^{\rm tot}$ 
contains field contributions. The third question is discussed in~\cite{sfc}, 
in which the dynamics of a hypothetical superfluid crystal is derived, where 
the equation of motion for the displacement vector $u_i$ is shown to be 
$\dot u_i=v^n_i+Y^D$, with $Y^D$ a dissipative term that accounts for 
diffusion of defects and interstitials~\cite{mpp}. So, if $Y^D$ is zero 
(especially true in equilibrium), the lattice points move with $v_n$. This 
remains valid also for superconductors, the dynamics of which reduces to 
that of a superfluid crystal in the limit $e\to0$ of Eq(\ref{1}). 

We now revisit Eq(\ref{4}): When deriving the constitutive relation for 
$j_s$, the information that it must be an invariant quantity was a 
consequence of asserting the consistency of Eq(\ref{3}). More prudently, 
this should be obtained as a result. So, the form of $j_s$ will be derived 
independently below, without the reference to Eq(\ref{3}), or even the input 
that $v_s/\eta$ is a velocity -- though the equivalent information of 
Eq(\ref{6}), the Josephson equation, is needed. As we shall see, this 
calculation will in addition provide further insights in, and understanding 
of, the properties of superconductors. We start the derivation by observing 
that the energy current ${\bf Q}$ contains the term ${\bf j}_s\mu$, then 
deduce the form of the total momentum density by the known symmetry of the 
energy stress 4-tensor, $\mbox{\boldmath$g$}^{\rm tot}={\bf Q}/c^2$; and 
finally, we obtain the explicit form of ${\bf j}_s$ via a Maxwell relation 
linking $\mbox{\boldmath$g$}^{\rm tot}$ to ${\bf j}_s$. 

The energy current $\bf Q$ is obtained by evaluating $\dot\varepsilon$ via 
Eq(\ref{5}), and requiring that it is given as a total divergence, 
$\dot\varepsilon=-\nabla\cdot{\bf Q}$. In the rest frame 
$\mbox{\boldmath$v$}_n=0$, and disregarding dissipative terms, we have $\dot 
s,\,\dot u_i=0$, hence only four terms remain:
$$\dot\varepsilon=\mu\dot\varrho +{\bf E}\cdot\dot{\bf D} +{\bf 
H}\cdot\dot{\bf B}+ {\bf j}_s\cdot\dot{\mbox{\boldmath$v$}}_s. $$
With $\dot{\mbox{\boldmath$v$}}_s$ given as in Eq(\ref{6}), the Maxwell 
equations and the continuity equation must contain the corresponding counter 
terms, 

\begin{eqnarray}\label{max1}
\dot{\bf B}=-c\nabla\times{\bf E},\\
\label{max2}\dot{\bf D}= c\nabla\times{\bf H}- e{\bf j}_s/m,\\
\dot\varrho+\nabla\cdot{\bf j}_s=0,\label{conti}
\end{eqnarray}
such that the energy current is ${\bf Q}= {\bf j}_s\mu+c{\bf E}\times{\bf 
H}$. The same calculation in a general frame $\mbox{\boldmath$v$}_n\not=0$ 
is tedious, but not more difficult. And of the new terms, all $\sim v_n$, 
the overwhelming one is that associated with the rest mass, $\varrho 
c^2\mbox{\boldmath$v$}_n$. So the total momentum density is 

\begin{equation}\label{g}
\mbox{\boldmath$g$}^{\rm tot}=\varrho\mbox{\boldmath$v$}_n+{\bf 
j}_s\mu/c^2+ {\bf E}\times{\bf H}/c, 
\end{equation}
while the neglected terms are again relativistic corrections, ones that are 
completely irrelevant in the present context. Now consider the 
Maxwell relation, 

\begin{equation}
\partial g_i^{\rm tot}/\partial j^s_j\ \vert_{v_n}=\partial v^s_j/\partial 
v^n_i\ \vert_{j_s} =(\mu/c^2)\delta_{ij}. \end{equation}
Evaluating the first expression via Eq(\ref{g}) we arrive at the third 
expression; and the second equal sign only allows a functional dependence as 
given in Eq(\ref{4}). This concludes the independent proof.  

The total momentum density, Eq(\ref{g}), may be written as 
$\mbox{\boldmath$g$}^{\rm tot}=(\varrho-\varrho_s)\mbox{\boldmath$v$}_n+ 
\varrho_s\mbox{\boldmath$v$}_s$ for $E=0$, implying a literal {\em 
nonclassical rotational inertia}~\cite{leg} for superconductors -- though it 
is of course much smaller than in He~II: Employing 
Eqs(\ref{3},\ref{4}), one finds in a superconducting cylinder of radius 
$R\gg\lambda$, that the superfluid velocity $v_s$ deviates from the normal 
one, $\mbox{\boldmath$v$}_n=\mbox{\boldmath$\Omega$}\times{\bf R}$, by the 
amount $\lambda (2\Omega+ \gamma B_{\rm 
ext})\exp[(r-R)/\lambda]$, ($r$ is the distance from the center, and $B_{\rm 
ext}$ the external field). Therefore, the fraction of $\varrho_s/\varrho$ of 
the total mass does not quite participate in the rotation of the rest, 
$(\varrho-\varrho_s)/\varrho$, and reduces (or enhances) the moment of 
inertia accordingly -- as a function of the external magnetic field.  

Eqs(\ref{max2},\ref{conti}) show that ${\bf j}_s\equiv\partial\varepsilon/
\partial\mbox{\boldmath$v$}_s$ transports both electric charge and 
mass in equilibrium, with a quotient given by the 
microscopic parameter $e/m$. As is clear from the derivation, this is 
intimately related to the form of the Josephson equation, and to the {\em 
nonclassical rotational inertia} -- or that $g^{\rm tot}\sim j_s$. If the 
Josephson equation, Eq(\ref{6}), were of the form $(\hbar/2m)\dot\varphi= 
-e\Phi/m$, the terms $\sim j_s$ in Eqs(\ref{conti},\ref{g}) (ie in the 
momentum density and mass current) would vanish; if it were 
$(\hbar/2m)\dot\varphi + \mu=0$, the persistent electric current, in 
Eq(\ref{max2}), is zero. The first is a hypothetical case of pure 
superfluidity in the electric charge, the second a case of pure 
superfluidity in the mass, such as realized in He~II. The specific sum 
in Eq(\ref{6}) characterizes a phase with condensed electrons, ones that 
carry mass and charge of the given ratio. 

All this is reminiscent of the relative broken symmetries in the 
superfluid phases of $^3$He~\cite{He} -- especially $^3$He-A$_1$, in which 
the condensation is in the up-spin population, of $^3$He particles that 
carry both spin and mass, with the ratio $\hbar/2m$. And the corresponding 
broken symmetry is a linear combination of spin and phase symmetry -- if one 
forgets the less obvious orbital one. The microscopic coefficient $\hbar/2m$ 
also characterizes the ratio between the equilibrium currents for 
mass and spin, and it enters the relevant Josephson equation~\cite{he} 
which -- as in Eq(\ref{6}) -- contains two thermodynamic conjugate 
variables, $\mu\equiv\partial\varepsilon/ \partial\varrho$ and 
$\mbox{\boldmath$\omega$}\equiv\partial\varepsilon/\partial {\bf s}$. This 
gives rise to a number of useful analogies that should be explored.

\end{multicols}

\begin{thebibliography}{99}
\bibitem[*]{email} e-mail: liu@itp.uni-hannover.de
\bibitem{London}F. London, {\em Superfluids} Vol I (New York, Wiley, 1950)
\bibitem{Cabr}B. Cabrera, Jap. J. Appl. Phys. {\bf 26-3} 1961 (1987), and 
references therein
\bibitem{laf}R.B. Laughlin Phys. Rev {\bf B 23} 5632 (1981); {\em The 
Quantum Hall Effect} eds R.E. Prange and S.M. Girvin, (Springer, New York, 
1987)
\bibitem{PWA}P.W. Anderson {\em Progress in Low Temperature Physics} ed. 
C.J. Gorter, (North Holland, Amsterdam, 1967)
\bibitem{Bra}R.M. Brady, J.Low Temp.Phys {\bf 49} 1 (1982)
\bibitem{baym}G. Baym {\em Frontier and Borderlines in Many Particle 
Physics} eds. E.A. Broglia and R. Schrieffer (Amsterdam, North Holland, 
1988) chap. 3
\bibitem{dG}P.G. de Gennes, {\em Superconductivity of Metals and Alloys} 
(Benjamin, New York, 1966) \S6-2
\bibitem{d}Particle Physics Booklet, Phys. Rev {\bf D 54} Part 2, 21 (1996)
\bibitem{HL} The Heaviside-Lorentz units, or the so-called {\em rational} 
units, employed in this paper renders the formulas and their manipulations 
much simpler: In this system, all four fields have the same dimension, 
square root of the energy density, (ie $\sqrt{\rm J/m^3}$ in MKS and 
$\sqrt{\rm erg/cm^3}$ in cgs,) sensibly with $H=B$ and $D=E$ in vacuum, and 
of similar magnitudes in ponderable media. As a result, the ubiquitous 
factors of $4\pi$, $\epsilon_0$ and $\mu_0$ --- the actual reason why we 
have to continually look up the formulas --- simply vanish. To revert to 
MKSA ($\hat E, \hat H$...), employ $$\hat H=H/\sqrt{\mu_o},\ \hat 
B=B\sqrt{\mu_o},\ \hat E=E/\sqrt{\epsilon_o}, \ \hat 
D=D\sqrt{\epsilon_o},$$
$$\hat\varrho_e=\varrho_e\sqrt{\epsilon_o},\quad\hat 
j_e=j_e\sqrt{\epsilon_o}. $$ 
To revert to the Gaussian system,  reduce all four fields, and increase all 
four sources, by the factor of $\sqrt{4\pi}$:\\ $ B^H=B^G/\sqrt{4\pi}$, 
(similarly for $D, H, E$); $\rho^H=\sqrt{4\pi}\rho^G$, (similarly for $j, P, 
M$). 
\bibitem{3cd}P. Kost\"adt and M. Liu, preprint
\bibitem{LL9}L.D. Landau  and  E.M. Lifshitz, Statistical Physics Part II 
(Pergamon 1989), \S 44
\bibitem{nurmax}M. Liu,  Phys. Rev. Lett. {\bf 70}, 3580 (1993); {\bf 74}, 
1884, (1995); {\bf 74}, 4535, (1995); {\bf 80}, 2937 (1998)
\bibitem{mpp}P.C. Martin, P. Parodi,  P.S. Pershan, Phys. Rev {\bf A6}, 240 
(1972)
\bibitem{kh}K. Henjes and M. Liu, Ann. Phys. {\bf 223}, 243 
(1993); Y-M. Jiang and M. Liu, Phys. Rev. Lett. {\bf 77}, 1043 (1996)
\bibitem{sfc}M. Liu, Phys. Rev {\bf B 18}, 1165 (1978)
\bibitem{leg}A.J. Leggett, in {\em Lecture Notes 
in Physics} 394, eds. M.J.R. Hoch and R.H. Lemmer (Springer, Berlin, 1991)
\bibitem{He}D. Vollhardt and P. W\"olfe,
 {\it The Superfluid Phases of Helium 3} (Taylor and Francis, London 1990). 
\bibitem{he} M. Liu and M.C. Cross, Phys. Rev. Lett. {\bf 
41}, 250 (1978); {\bf 43}, 296 (1979); M. Liu, Phys. Rev. Lett. {\bf 43}, 
1740 (1979); Physica {\bf 109\& 110B}, 1615 (1982)
\end{thebibliography}
\end{document}